\makeatletter
\declare@file@substitution{revtex4-1.cls}{revtex4-2.cls}
\makeatother

\documentclass[twocolumn,times,tighten]{aastex63}

\usepackage{color, enumitem, hyperref, verbatim, amsmath, amstext, mathrsfs, hypcap, afterpage, marginnote, makecell, subfigure}
\usepackage{multirow}
\usepackage{booktabs}
\usepackage{natbib}
\usepackage{comment}
\usepackage{graphicx}	
\usepackage{amsmath}	
\usepackage{amssymb}	
\usepackage{lineno}

\newcommand*{\kh}{}
\newcommand*{\toreferee}{}
\newcommand*{\torefereetwo}{} 
\graphicspath{{./}{figures/}}

\received{\today}

\shorttitle{}

\shortauthors{Ho, Yuen, \& Lazarian}

\graphicspath{{./}{figures/}}

\begin{document}

\title{ The stable "Unstable Natural Media" due to the presence of turbulence
\footnote{Drafted on \today}}

\author{Ka Wai Ho}
\email{kho33@wisc.edu}
\affiliation{Department of Astronomy, University of Wisconsin-Madison, USA}
\affiliation{Theoretical Division, Los Alamos National Laboratory, Los Alamos, NM 87545, USA}

\author{Ka Ho Yuen}
\affiliation{Theoretical Division, Los Alamos National Laboratory, Los Alamos, NM 87545, USA}

\author{Alex Lazarian}
\affiliation{Department of Astronomy, University of Wisconsin-Madison, USA}



\begin{abstract}
{\toreferee The term "unstable neutral media" (UNM) has traditionally been used to describe the transient phase formed between the warm and cold neutral hydrogen (HI) phases.} However, recent observations suggest that the UNM phase not only has a significantly longer-than-expected lifetime but also occupies at least 20\% of both the volume and mass fraction of HI. In this paper, we argue that the existence and dominance of the UNM can be explained by the presence of strong turbulence using an energy balance argument. The mass fraction of UNM is directly proportional to the turbulent velocity dispersion $\sigma_v$: mass fraction of UNM $\propto \sigma_v^{\frac{2n}{1+n}}$, where $n$ is the absolute value of the adiabatic index in the unstable phase. We discuss the implications of long-lived unstable thermal phases on ISM physics, including cold dense filament formation, cosmic ray acceleration, and measurement of galactic foreground statistics.
\end{abstract}


\section{Introduction} \label{sec:intro}
The interstellar medium (ISM) is a complex mixture of gas, dust, and cosmic rays that plays a crucial role in the evolution of galaxies (\citealt{1977ApJ...218..148M}, see also \citealt{MO07,filament_review}). The ISM can be divided into several phases, including the warm neutral medium (WNM), cold neutral medium (CNM), and unstable neutral medium (UNM, \citealt{2003ApJ...586.1067H}). The WNM is characterized by temperatures of several thousand Kelvin and low densities, while the CNM has temperatures around 100 K and higher densities \citep{2003ApJ...587..278W}. The UNM is a dynamic and transitional phase that lies between the WNM and CNM and is susceptible to rapid changes in temperature and density \citep{1965ApJ...142..531F}.

{\kh A widely-held assumption in the ISM community \citep{1977ApJ...218..148M,filament_review} suggests that the unstable phase, as its name implies, is thermally unstable and has a relatively short lifetime. This argument is based on the fact that when the cooling timescale, influenced by various processes including radiative cooling, chemical reactions, and turbulence \citep{2007IAUS..237..306N}, is balanced by the turbulence timescale. A rough estimation shows that the UNM lifetime is on the order of $O(1Myr)$. If this were the case, since the mass of the galaxy is conserved, the UNM fraction in our Milky Way should have declined very quickly, given that the lifetime of the galaxy is significantly longer than that of the UNM. However, numerical simulations suggest otherwise \citep{2018ApJ...853..173K,2020MNRAS.492.1465S,2020MNRAS.497.4196S}. Indeed, earlier literature with weak turbulence suggests that the fraction of UNM is on the order of a few percent \citep{2000ApJ...540..271V, 2005A&A...433....1A,2012MNRAS.424.2599C}. However, it has been observationally suggested \citep{2009ARA&A..47...27K,2018ApJS..238...14M} that UNM occupies {\bf at least} 20\% of the total fraction \citep{2018ApJS..238...14M,2023ARA&A..61...19M}, with some reports of up to 40\% \citep{2018A&A...619A..58K} {\toreferee with an uncertaincy of 20\%}, suggesting that there are some sort of 'heat source' continually elevating the CNM to warmer phases.}


One of the leading hypotheses is that the presence of turbulence acts like a heat source, maintaining the high fraction of UNM observed. Recent studies have shown that turbulence can significantly impact the fraction and stability of the UNM \citep{2005A&A...433....1A,2007A&A...465..431H,2007A&A...465..445H}. Moreover, new evidence suggests that turbulence plays an important role in shaping the power spectrum \citep{spectrum} and anisotropy \citep{instability} of the multiphase ISM. In fact, earlier literature \citep{2003ApJ...589L..77C} suggests that turbulence can transport heat much more effectively than the native thermal conduction rate, and in the case of the ISM, it is a few orders of magnitude stronger than its thermal counterpart. {\toreferee Despite there are several studies showing a strong connection between the UNM fraction and the level of turbulence driving \citep{2000ApJ...540..271V,2002ApJ...577..768S,2005A&A...433....1A,2007A&A...465..431H,2007A&A...465..445H}}, a quantitative analysis of how this leads to a large fraction of UNM is still {\toreferee in question}.

In this paper, we discuss how turbulence can extend the lifetime of the unstable phase, making it appear stable compared to the radiative cooling timescale. In Section \ref{sec:Theory}, we briefly review the relevant timescales for multiphase physics. In Section \ref{sec:eb}, we utilize the energy argument to explain why turbulence can significantly enhance the UNM fraction as observed in the sky. In Section \ref{sec:num}, we discuss our numerical results. In Section \ref{sec:conclusion}, we summarize and conclude our paper.

\section{Review of timescale argument in formation and stability of unstable phase}
\label{sec:Theory}

\subsection{Important time scale and length scale in Multi-phase ISM}
When discussing the physics of multi-phase ISM, the cooling effect is considered as one of the dominating process in studying the dynamics of ISM. To demonstrate that, one of the classical argument is coming from the timescale comparison. The multi-phase ISM can be treated as a fluid, and for that, the dynamical timescale $\tau_d$ determines the importance of fluid motion. For cooling effect, we can define a cooling time scale $\tau_c$. They can be defined as :
\begin{equation}
\begin{aligned}
\tau_c = \frac{E}{\dot{E_{c}}},
\\
\tau_d = \frac{L}{c_s},
\end{aligned}
\end{equation}
where $E,\dot{E_c},L,c_s$ are the internal energy of the fluid element, cooling rate, length scale of the ISM and speed of sound. The timescales can be interpreted as the time required for a process become important. The shorter one is more important than a longer one. 

Assuming the idea gas law and plugging in the {\toreferee typical} ISM parameters, we can then estimate the two timescales. For example, assuming the ISM temperature $T=1500K$, $L=100pc$, and $\dot{E_c} = 2\times 10^{-26} erg s^{-1}$ and density $n_H = 2 cm^{-3}$, we arrive with an order of magnitude estimate of $\tau_{d} \sim O(10 Myr)$ and $\tau_{c} \sim O(Myr)$, indicating that the cooling finishes almost instantly after the heat exchange due to fluid motion. We neglect the effect from thermal conduction as the corresponding characteristic length scale is at the order of $10^{-2}$ pc, meaning that the natural thermal conduction only plays a minor role on the scale of ISM, consistent with the estimate in \cite{2003ApJ...589L..77C}.

In this interpretation, we observe that the internal energy of fluid element ultimately determined by the cooling function. The multiphase gas will be stabilize very quickly roughly at the order of cooling timescale, where a static fraction of warm and cold gas mixture are formed without much mixing, which is the scenario outlined as early as \cite{2003ApJ...587..278W}.

\begin{figure}
\label{fig:illus}
\centering
\includegraphics[width=0.26\paperheight]{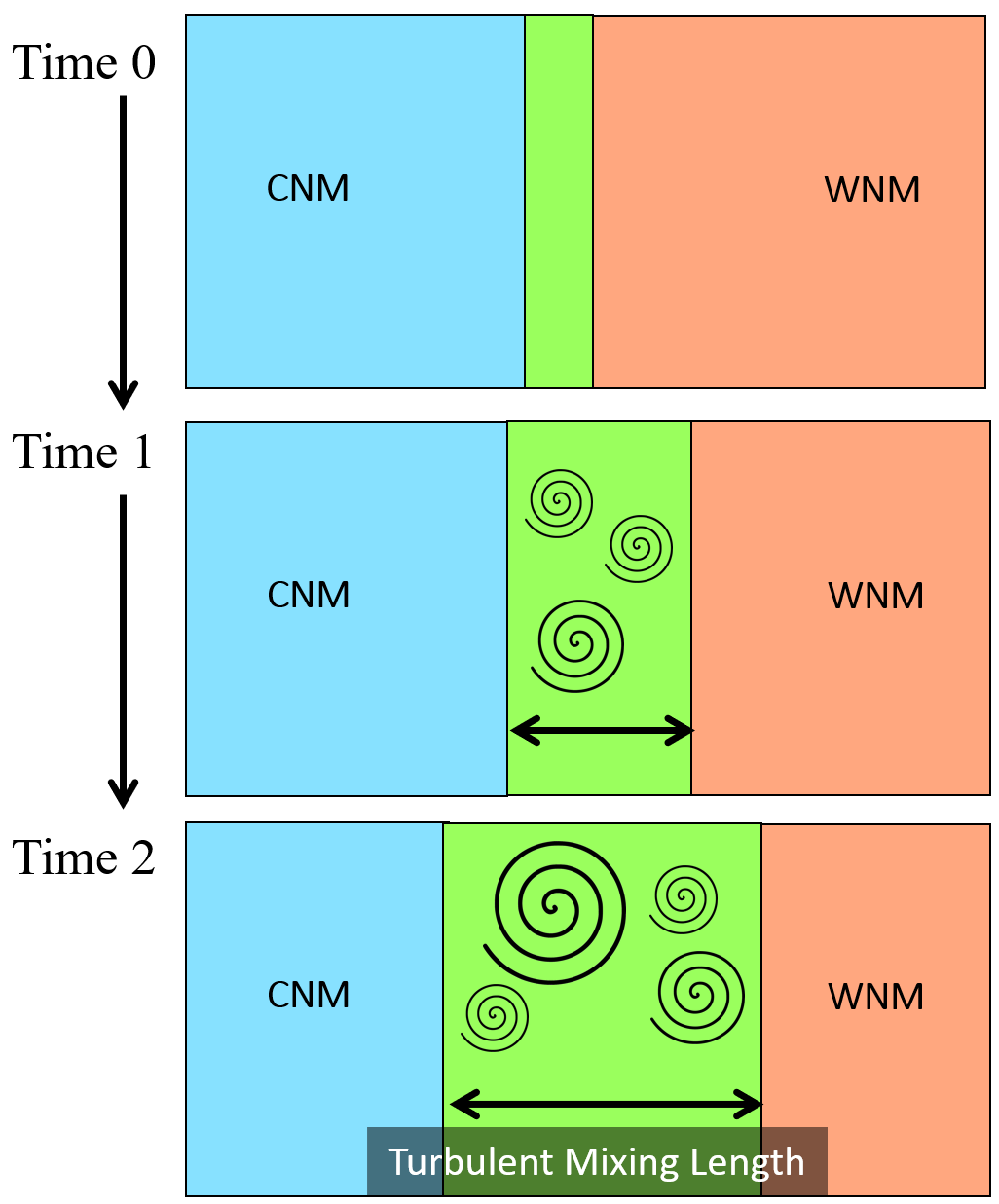}
\caption{Illustration of Cold-Heat End Model, The formation of UNM under temperature gradient and turbulence mixing.}
\end{figure}

\subsection{Turbulent mixing effect}
However, observational study shows that the multi-phase ISM is turbulent (See, e.g. \citealt{spectrum}) and one should consider the effect arising from turbulent motion. Prior simulations observed that a substantial amount of unstable phases are produced when turbulence is generated, but the actual fraction of unstable phases are not in agreement, despite a rough range of estimate of $20-28\%$ is observed across different simulations\citep{2020MNRAS.492.1465S,2020MNRAS.497.4196S,2023ApJ...949L...5F}, but some of other simulations has the unstable phases up to $60\%$ \citep{2017NJPh...19f5003K,2018PhRvL.121b1104K}. Similar variation is also raised from the observational side \citep{2018ApJS..238...14M,2020A&A...639A..26K,2023ARA&A..61...19M}. 

For a fully turbulent medium, turbulence mixing effect plays an important role in transferring heat of the fluid, as turbulence heat transport known to be more efficient than natural thermal transport in the case of ISM environment\citep{2003ApJ...589L..77C}. To quantify the turbulence mixing, we follow the argument from last section and consider its timescale, which defined as :
\begin{equation}
\begin{aligned}
\tau_{t} = \frac{l}{v},
\end{aligned}
\end{equation}
where $l,v$ are the eddies scale and length scale {\it at the size of unstable phase}. One could link the two observables via \citeauthor{K41} (\citeyear{K41}) scaling $v \propto l^{1/3} $ and arrive with:
\begin{equation}
\begin{aligned}
\tau_{t} \sim \tau_0 \left(\frac{l}{L}\right)^{2/3},
\end{aligned}
\end{equation}
where $\tau_0 = L/V_0$ denoting the turbulence mixing time at timescale given the velocity $V_0$ at scale L. One could see that the turbulence mixing effect is scale dependent and decreasing in smaller scale. We consider at a smaller length scale $l_t$, $\tau_t$ would decrease down to the magnitude of $\tau_c$. To estimate the value, we assume that at the unstable phase it is mildly trans-Alfvenic, i.e. $V_0 \sim c_s$, and the timescale estimate gives:
 \begin{equation}
\begin{aligned}
\left(\frac{\tau_c}{\tau_d}\right)^{3/2} \sim \left(\frac{l_t}{L}\right).
\end{aligned}
\end{equation}
For the case of $\tau_c/\tau_d \sim 0.1$ that we adopted earlier ($\tau_d \sim 10Myr, \tau_c\sim 1 Myr$), $l_t$ is about $3 pc$, meaning that at this scale, the effect of turbulence mixing effect play a comparable role with cooling effect. Fig.\ref{fig:illus} shows a cartoon on how the turbulence mixing actually works: When turbulence is in play, there is an intermediate layer with the width be the turbulence mixing length $l_t$ in which the energy is supported by turbulence free energy. While the two stable phases have no heat loss, fluid elements transported into the mixing layer are subjected to imbalanced heating and cooling which is reflected by the negative adiabatic index as seen in multiphase numerical simulation, which we shall discuss in the next section. 

\section{Energy balance argument}
\label{sec:eb}

{ 

From \S \ref{sec:Theory}, we observe that the turbulence mixing effect dominates over thermal conduction in terms of heat transport. However, we do not see why the argument in the previous section could explain the potential very long lifetime and large fraction of UNM. The reason is because the modern MHD turbulence theory \citep{GS95,2001ApJ...554.1175M,CL03}  assumes that the only energy sink for the plasma modes are dissipation (from ion-neutral damping in the case of ISM, see \cite{2004ApJ...614..757Y,2015ApJ...810...44X}). However, in the case of multiphase media, there is an additional energy sink that allows the turbulence to behave differently during phase transition: the abnormal {\toreferee equilibrium} $P\propto \rho^{-n} (n>0) $ unstable phase equation of state (EoS). While we will postpone the full analysis to the upcoming paper (Yuen et al. in prep) in exploring the quantitative values of $n$ as expected for a given turbulence model, it is crucial to consider how turbulence energy transfer is changed when having the abnormal EoS. {\toreferee In this section, we focus on the case of hydrodynamic case, while we will discuss the case with magnetic field in Appendix \ref{ap:A}. }

Let us consider the scale of UNM with thickness of $\sim 1pc$, which is evidently larger than the ion-neutral decoupling scale \citep{2010ApJ...718..905L,2010ApJ...720..603H,2011MPLA...26..235H,2015ApJ...810...44X}. The energy equation can be written as ($\epsilon = \rho v^2/2$):
\begin{equation}
\frac{\partial \epsilon}{\partial t} + \nabla\cdot ((2\epsilon + P){\bf v}) = \rho \Lambda - \rho^2 T^{1/2}\Gamma e^{-184/T}
\label{eq:E_master}
\end{equation}
where $ \rho \Lambda$ is the radiative heating function and $\rho^2 T^{1/2}\Gamma e^{-184/T}$ is the radiative cooling function. Both $\Lambda$ and $\Gamma$ can be approximated as roughly constants ({\toreferee See \citealt{1983ApJ...270..511Z,2002ApJ...564L..97K} and also numerical realization \citealt{2017ApJ...834...25K,2018PhRvL.121b1104K}}) in the length scales that we are considering. The rough variation of the heating and cooling functions as the temperature changes is shown in {Fig \ref{fig:phase_diagram}}.  A general trend for the heating and cooling functions are: in the case of WNM and CNM they are roughly balanced, and an adiabatic EoS is enforced. For UNM, the cooling function is significantly stronger than that of the heating function. Notice that we are assuming that our mixture of gases are in the regime 3 of \cite{2003ApJ...587..278W}, i.e., the mean density (so as pressure) lies in between the maximum warm phase density and the minimum of cold phase density. In our discussion, we can therefore combine both CNM and WNM together, while separating that of UNM for dedicated analysis. Our ultimate goal is to obtain the following parameter \footnote{Here, one has to be careful that $\rho(T)$ us actually $\rho f(T)$, where $f$ is the distribution of phases in temperature \citep{2023ApJ...946....3K}}. We discuss a more first-principle method in Appendix \ref{ap:B}. }:
\begin{equation}
\chi_{UNM,mass} = \frac{\int_{min(T): \Lambda= \rho T^{1/2}\Gamma e^{-184/T}}^{max(T): \Lambda= \rho T^{1/2}\Gamma e^{-184/T}} dT \rho(T)}{\int dT \rho(T)}
\label{eq:7}
\end{equation}
One of the major principles for us to move forward is that all three phases shared the same Kolmogorov scaling \citep{spectrum}. i.e  their $\delta v$ strictly obeys $\delta v \propto l_\perp^{1/3}$. Noticing that the solution of the integral in Eq.\ref{eq:7} is Equation of State dependent. At the moment, we will assume an isotropic scaling law to provide a first order estimate and defer the anisotropy argument in later publication.\linebreak

From \cite{spectrum} we observe that the energy transfer rate $\frac{\partial \epsilon}{\partial t}$ is roughly constant since apparently a Kolmogorov-like spectrum is maintained over the scales of transitions. We can proceed with Eq.\ref{eq:E_master} by expressing $P$ and $T$ via the polytropic EoS with negative adiabatic index and ideal gas law, respectively:
\begin{equation}
\begin{aligned}
P &= C\rho^{-n} \quad (n>0)\\
T &= \frac{P}{\rho k_B} = \frac{C}{k_B} \rho^{-n-1}
\end{aligned}
\end{equation}
for some constant C. {\torefereetwo The expected value of $n>0$ in HI emission observation in the local universe {\toreferee (See, e.g. \citealt{2009ARA&A..47...27K}). We want to remind the readers that, the observations on $n$ is not direct and there are only a few available data points from literature. In the case of high-redshift universe, it is shown theorerically that $n<0$ is possible due to insufficient cooling line strength \citep{2016ApJ...822...83B}.
}}. We first approximate the range of values of $T$ in relation to $n$:
\begin{equation}
 C' T^{\frac{n-1}{2(n+1)}}  \approx  e^{184/T}
\end{equation}
for some constant $C'$, denoting these two solutions as $T_{1,2}$ that correspond to the cold and warm gas density, respectively. Then Eq.\ref{eq:7} can be written as \footnote{When the system is isobaric ($n=0$), the current integral gives a logarithmic form. 
{\torefereetwo We recognize that the current analytical construction is not suitable in describing the case when the phase diagram does not exhibit a range of negative polytropic equation of state. This effect is particularly prominent when (1) there is no magnetic field, (2) the metallicity is too low. The latter case only happens in high redshift ISM \citep{2016ApJ...822...83B}. However, the former case deserves further physical discussion, which we will explain in the Appendix \ref{ap:A}. }}:
\begin{equation}
\begin{aligned}
m_{UNM} &\propto \int_{T_1}^{T_2} dT (\frac{C}{k_B})^{\frac{1}{n+1}} T^{\frac{-1}{n+1}}\\
&= (\frac{C}{k_B})^{\frac{1}{n+1}} \frac{n}{n+1} (T_2^{\frac{n}{n+1}} -T_1^{\frac{n}{n+1}})
\end{aligned}
\end{equation}
where from observation we know $T_2 \gg T_1$ ($T_1 \approx 200K, T_2 \approx 5000K$), therefore the $T_1$ part can be safely ignored in our order-of-magnitude estimates.

What remains is to relate $T_{1,2}$ with turbulence velocity $v$. Assuming all the turbulence energy went into the support of the negative adiabatic EoS:
\begin{equation}
\frac{2\rho v^3}{l} \sim  \frac{C\rho^{-n}v}{l} 
\label{eq:10}
\end{equation}
Taking Kolmogorov scaling {\toreferee for velocity fluctuations}\footnote{We want to emphasize that the velocity scaling (thin channel velocity scaling as outlined in \citealt{LP00} and the density scaling (column density scaling) are two different things. For the latter, which is more observationally measured, the phases appear to have different scaling laws. See \cite{2018ApJ...856..136P} for the density scaling, while \cite{spectrum} for the velocity channel scaling.}, we have $v^3/l \approx \text{const}$ 
\begin{equation}
\begin{aligned}
v^{-2} &\propto l^{-2/3} \propto   \rho^{1+n}  \propto T^{-1}\\
\rightarrow T^{1/2} &\sim v
\end{aligned}
\label{eq:11}
\end{equation}
The mass fraction of UNM is given by ($n>0$, see Footnote 1):
\begin{equation}
\chi_{UNM,mass}  = \text{const}\times v^{\frac{2n}{1+n}}
\label{eq:prediction}
\end{equation}
For typical $n$ value of $0.3-0.4$ (i.e. $P\propto \rho^{-0.3} \text{ to } \rho^{-0.4}$), the mass fraction of UNM is about $v^{0.46} \text{ to}\:v^{0.57}$. {\torefereetwo We remind the readers that the current study focuses on the case neglecting the importance of magnetic field. In Appendix \ref{ap:A} we discuss the case by including the role of magnetic field.}

\begin{table}
\centering
\begin{tabular}{@{}cc@{}}
\toprule
Simulation & \( \sigma_v\:[km\:s^{-1}]\)\\\midrule
M0         & 0.30                       \\
M1         & 0.39                       \\
M2         & 0.75                       \\
M3         & 1.53                       \\
M4         & 2.65                       \\
M5         & 5.59                       \\
M6         & 8.15                       \\ \bottomrule
\end{tabular}
\caption{ {\toreferee Simulation parameters, where velocity dispersion $\sigma_v$ (in km/s). Resolution $N^3=512^3$, Length Scale $L=100pc$, and mean number density of atomic hydrogen $n_H=3 cm^{-3}$. The mean magnetic field strength $B = 0.5 \mu G$}.}
\label{tab:sim}
\end{table}

\section{Numerical Method \& Result}
\label{sec:num}
\subsection{Numerical Simulations}
To compare the result in \S \ref{sec:Theory}, we use the 3D MHD multi-phase simulations generated from the MHD code Athena++, which were also being used in \cite{VDA} and \cite{instability}. For the initial state, we set up a 3D periodic turbulence box with the length of 100 pc and we are assuming the fluid represents the bulk neutral hydrogen in the interstellar media. We adopt the realistic cooling and heating function proposed by \cite{2002ApJ...564L..97K}. The simulation was originally constant in density and was driven via spectral velocity perturbation in the Fourier space. 

We set up a few simulations with the conditions similar to the realistic multi-phase neutral hydrogen gases with mass/volume fractions being consistent with observations.  We shall define the gas as the cold phase when the temperature of the gas is below 200K while those above 5500K as to warm phase, while the gas in between is the unstable phase. {\toreferee At around 50 Myr the turbulence box has produced a realistic multi-phase medium, which the simulation parameters are listed in Table. \ref{tab:sim}}. For our current case study, the cooling lengthscale is roughly $\sim L_{inj}/10$. Therefore, simulations at resolution of $512^3$ are sufficient for the current studies.

\begin{figure*}
\includegraphics[width=0.98\textwidth]{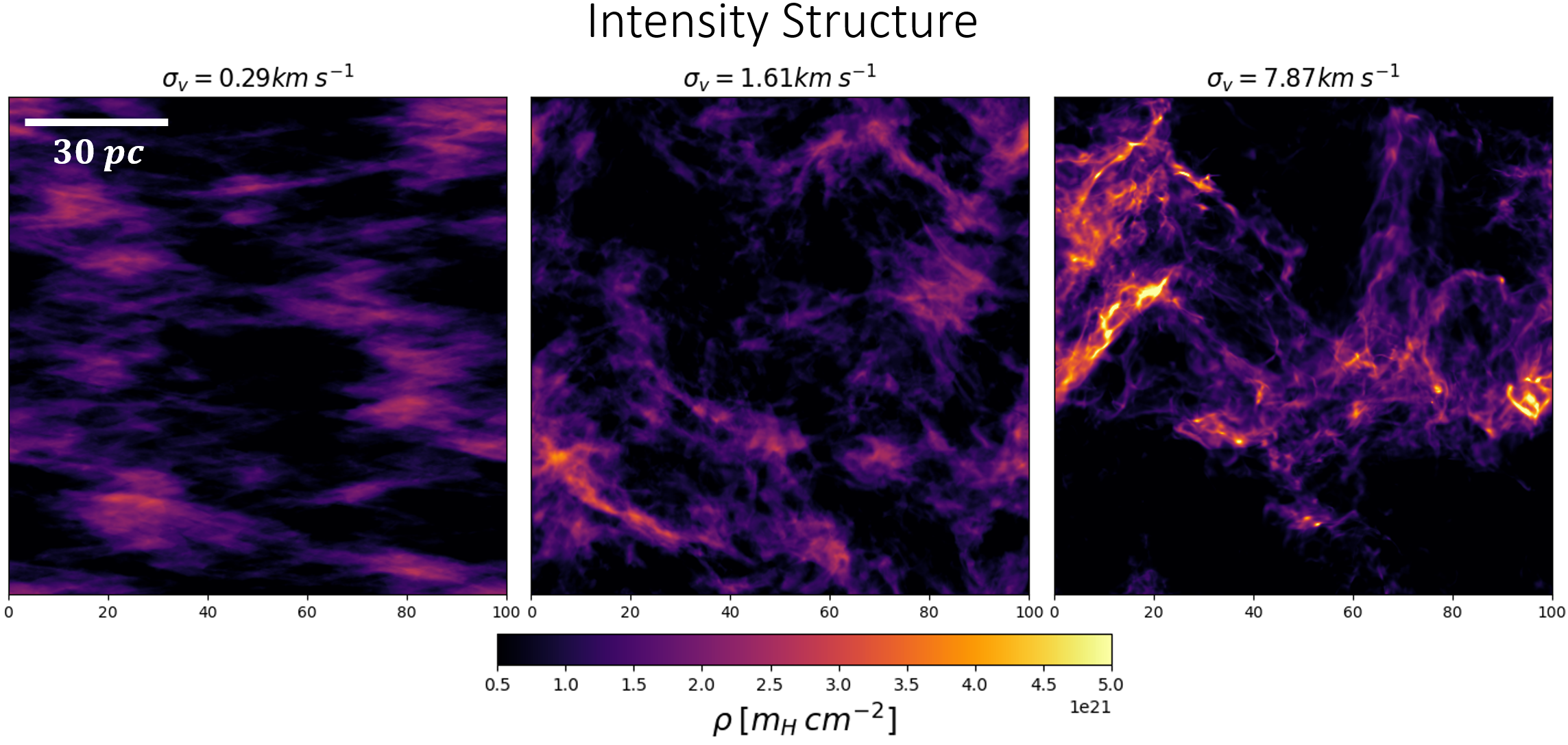}
\caption{\label{fig:intensity} {\toreferee Three panels showing the structure of multiphase HI density projection when we vary the velocity dispersion.}}
\end{figure*}

\begin{figure}
\includegraphics[width=0.30\paperheight]{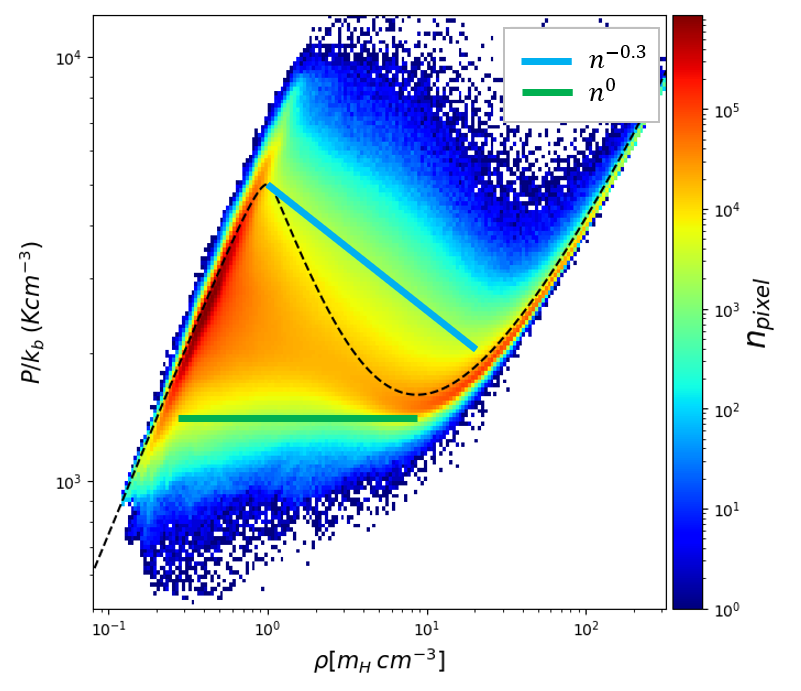}
\caption{\label{fig:phase_diagram} {\torefereetwo A plot shows the phase diagram (Simulation used: M5) of the simulations. The dashed black curve represents the equilibrium curve (i.e.\ heating = cooling in the energy evolution [R.H.S.\ of Eq.5=0]). We compare two power-law relations: $n^0$ (solid green line) representing isobaric behavior, and $n^{-0.3}$ (solid blue line). The gas distribution shows a negative adiabatic index ($n < 0$) in the unstable neutral medium (UNM) region, particularly evident at higher pressures, indicating non-isobaric behavior.}}
\label{fig:phase_diagram}
\end{figure}

\begin{figure*}
\centering
\includegraphics[width=0.99\textwidth]{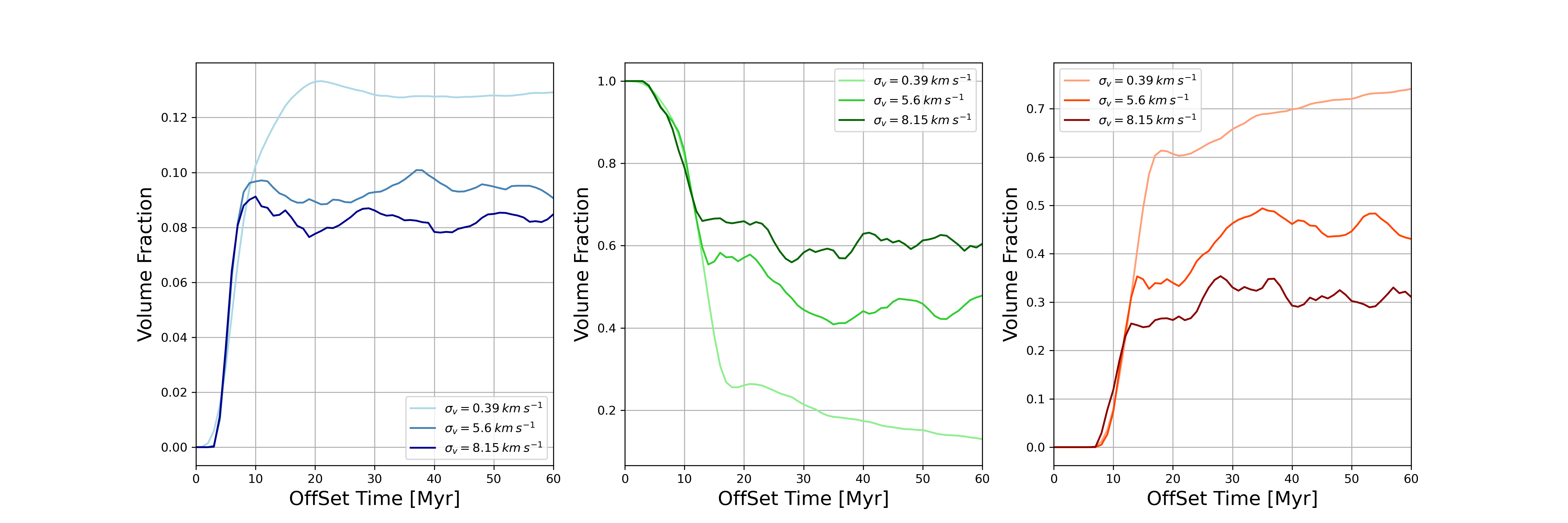}
\caption{\label{fig:v_frac} {\toreferee A set of figures showing how the volume fraction of the three phases (left: cold, medium: unstable, right: warm) vary across offset time for three different cases of turbulent velocity dispersion. The offset time defined as the time when thermal instability happens.}}
\end{figure*}

Fig. \ref{fig:intensity} shows how does the intensity structure (density projected along of sight) looks like in three cases of multiphase media with different injection velocity dispersion. We can see visually from Fig.\ref{fig:intensity} that the increase of injection velocity have apparent visual effect on the distribution of density features in the simulation domain, very similar to the situation that outlined in \cite{GA}: When the injection velocity is larger, the effective sonic Mach number is also larger, and as a result, thin features can be formed very easily under strong turbulence compression. 

We also present the phase diagrams of these simulations as in Fig. \ref{fig:phase_diagram}. We can see that stronger turbulence makes the thermal curve much more steepened, which is a well-known result from the community \citep{2017NJPh...19f5003K,2018ApJ...853..173K}. By observation, the numerical value of the adiabatic index $n$ in the unstable phase is about 0.3. We expect that the numerical value of $n$ to be shallower than the equilibrium curve (dashed line of Eq.\ref{fig:phase_diagram}) due to strong thermal instability.

\subsection{Numerical Verification of the fraction estimation } 

How does the turbulent velocity affects the fraction of the three phases?  In Fig.\ref{fig:v_frac} we show how the fraction of three phases vary as a function of offset time for three different cases of turbulent velocity.  We observe a few different things that are very apparent from these simulations: (i) when the injection velocity increases, the volume fraction of unstable phase increases. (ii) Both volume fractions of cold and warm phases decrease as the turbulent velocity increases, (iii) the volume fraction of cold phase enters the equilibrium earlier when the turbulence velocity amplitude is larger, (iv) all three phases have entered equilibrium stage for an extended amount of lifetime. These qualitative facts suggest that the increases presence of turbulence allow the originally thermally unstable phase becomes particularly stable in our study.

How does the fraction  stability be achieved? Earlier proposal \citep{2003ApJ...587..278W,MO07} suggests that materials are cycling between the cold and warm phases, which appears to make the unstable phase be in large fraction despite the lifetime of the thermally unstable materials are dynamically short. Fig.\ref{fig:v_frac} seems to suggest that both fractions of cold and warm phases are reduced as the turbulence strength increases, implying that the material cycling argument could work. Indeed, the increased presence of thermally unstable phase \citep{instability} imposes additional forces to cold phases. When UNM fraction increases, the thermal instability creates more CNM, creating an equilibrium in between. 

\begin{figure}
\includegraphics[width=0.48\textwidth]{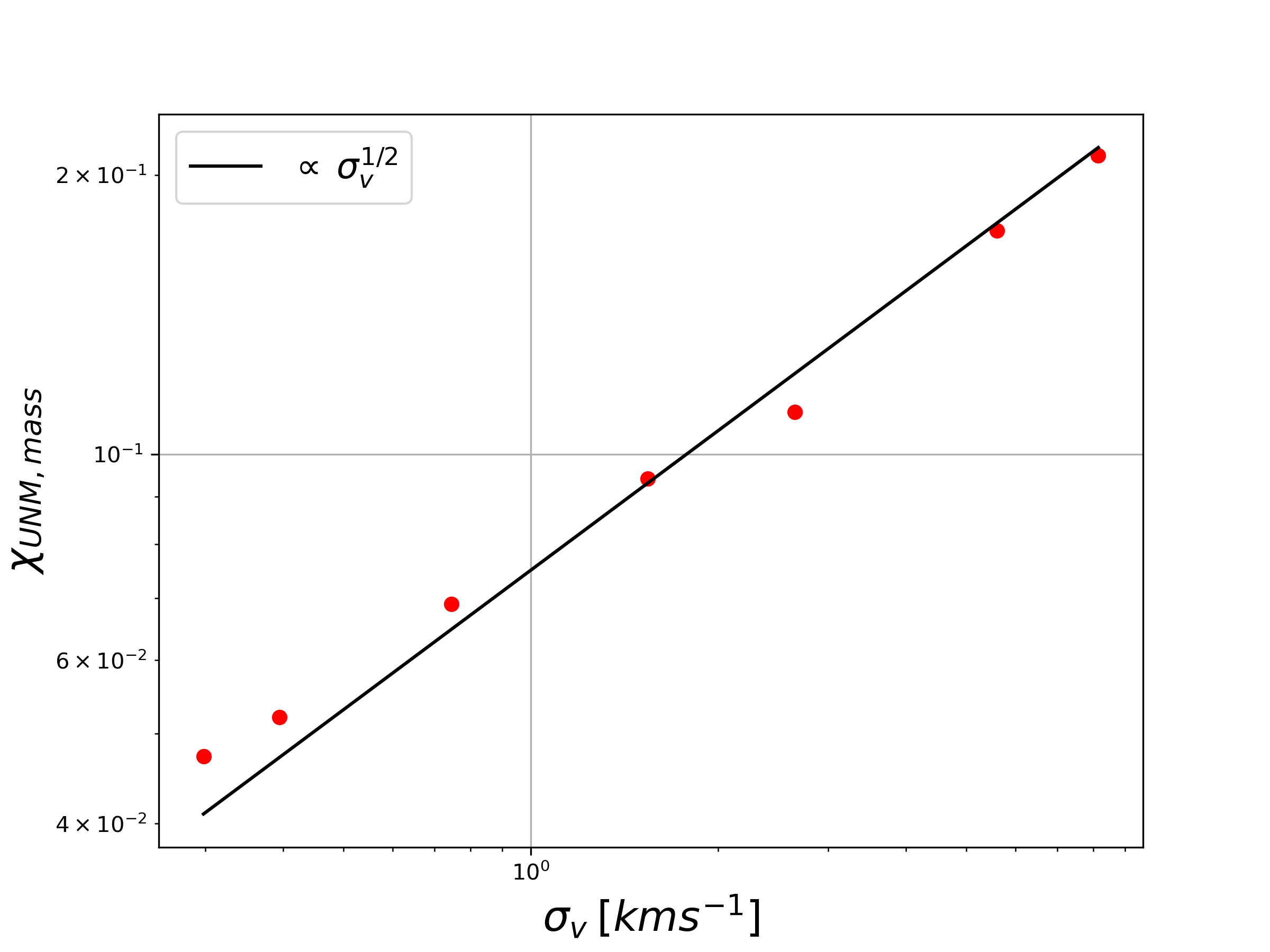}
\caption{\label{fig:v_frac_scaling} {\toreferee A figure showing how the time averaged mass fraction of unstable phase varies as a function of velocity dispersion. For each data point, it takes the average of 50 Myr after the saturation of turbulence.}}
\end{figure}

From the qualitative perspective we understand that turbulence fueled the stability of thermally unstable phase, but did our quantitative prediction (Eq.\ref{eq:prediction}) be realized in the simulation? In Fig. \ref{fig:v_frac_scaling} we show the time averaged { mass fraction} of the UNM as a function of $\sigma_v$, velocity dispersion, which is a direct proxy of $v$. Notice that in a simulation with given initial density and pre-defined heating and cooling functions, we already know approximately the mean density of the unstable phase. Fig.\ref{fig:v_frac_scaling} shows that the variation of { mass fraction} for the unstable phase has a power-law relation to the {injection velocity}. In our case the relation is roughly $\chi_{UNM} \propto v^{1/2}$,{\torefereetwo meaning that $n\approx 1/3$, matching the slopes in the phase diagram we see from Fig.\ref{fig:phase_diagram} if magnetic fluctuation is present (See Appendix \ref{ap:A}). }

\section{Discussion and Conclusion} \label{sec:conclusion}

In the previous paper of our series \citep{instability}, we revealed that the presence of negative adiabatic index will create additional, short-range gravity-like pressure term that confines the cold neutral media, i.e. long filaments (\citealt{2015PhRvL.115x1302C,2019ApJ...878..157X}, see \citealt{filament_review} for a review) that are longer than the threshold of the thermal instability are not stable. In the current paper we revealed that it is turbulence that supports this action, and a stronger level of turbulence will actually make the instability strong, instead of weaker. In other words, the density map of multiphase media under the combined action of heating and cooling functions will appear to be much more scattered, which agrees with our observation in Fig.\ref{fig:intensity}. We want to emphasize that, isothermal simulations with plasma conditions similar to that in our simulated cold neutral media {\bf will not} be that scattered, which can be readily seen in some state-of-the-art numerical simulations (e.g. \citealt{2021NatAs...5..365F}). It is a combined action of the turbulence and phase effect that drives both the dynamics \citep{instability} and the unstable fraction, which is presented in the current paper, to be very different from the isothermal counterpart.

{\toreferee Earlier publications have also computed the phase fractions in different numerical setting. Some of which includes incompressible velocity strain in 2D HD turbulence \citep{2005A&A...433....1A}, colliding flow \citep{2012MNRAS.424.2599C}, modification of heating rate \citep{2018ApJ...862...55H}, and chemical network modification to heating and cooling of multiphase gases \citep{2023ApJ...946....3K}. We want to emphasize that our simulations, which is a magnetized turbulence simulation in a triply periodic box with a very simplified form of heating and cooling \citep{2002ApJ...564L..97K}, still produce a scaling between the gas fraction to the turbulence velocity (Fig.\ref{fig:v_frac_scaling}). In this setup, the closest study that we can compare with is \cite{2005A&A...433....1A}, where they point out that the unstable phase fraction is a constant of velocity strain which the latter is naturally proportional to turbulence strength, we see a strong variation on unstable phase fraction relative to turbulence velocity fluctuations. This discrepancy may arise from the fundamental difference in dimensionality between our 3D simulations and their 2D approach, as the two cases exhibit fundamentally different turbulence cascade properties. } 

{\toreferee \noindent {\it Impact and caveat of our studies. --- } We want to emphasize that, in general, the phase fraction is also a function of the other parameters defined by the galactic environments. Our studies provide a theoretically simple and numerically precise way in correlating the unstable phase fraction to the turbulence strength. However, we emphasize that the way that we are modelling {\it assumes} the an ISM volume that contains saturated turbulence, which is a reasonable assumption within $\sim 100pc$ \citep{2010ApJ...710..853C}. } 

{\toreferee  In addition, the ionization fraction in multiphase turbulence changes significantly from $\chi_e = n_e/n_H \sim 10^{-1}$ in warm phase to $\chi_e \sim 10^{-3}$ in cold phases \citep{2023ApJS..264...10K}, causing potential ambipolar diffusion \citep{2015ApJ...810...44X} among phases. The exploration of this effect require a more realistic chemistry network \citep{2023ApJS..264...10K} and non-ideal MHD terms, which will be considered in later publications.} 





{\toreferee \noindent {\it Impact to ISM modelling.---}} We also want to emphasize that the micro-physics of unstable media is very important in further quantifying the dynamics of ISM (\citealt{2023ApJ...949L...5F,2024AAS...24421502K}). One of the most important development in the studies of ISM is the inclusion of low energy (1-200GeV) cosmic rays into the multiphase ISM system \citep{2024arXiv240307976H,2024AAS...24412302G}. It is observed that the compressive part of the turbulence energy, presumably most effectively generated by thermal instability, is taken away from the ISM in energizing (heating) the cosmic ray, leading to a runaway energy growth of cosmic ray energy. In the context of MHD turbulence theory, it is the fast modes that energize the cosmic rays most efficiently at that energy range \citep{2002PhRvL..89B1102Y}. Numerically it is reported that fast modes are dominant in the multiphase ISM \citep{2024AAS...24411402B}. We want to stress the fact that the dominance of fast mode also significantly modify the statistics of turbulence, particularly in the form of intermittancy \citep{2021ApJ...911...53H}, in which it accelerates cosmic rays \citep{2022MNRAS.514..657K} and modifies the observed polarization \citep{2024arXiv240517985P}. Furthermore, it appears to be sufficient in modifying the statistics of {\toreferee galactic} E/B modes on the sky (\citealt{2018PhRvL.121b1104K,2024AAS...24421502K,ho...2024}).


{\toreferee \noindent {\it Impact to cosmic ray transport.---}} The microphysics that we are investigating in this paper is incomplete. Despite we know that the phase change can lead to the spatial deformation of CNM \citep{instability}, and turbulence can also fueled up the phase changes (the current paper), we still do not know how the heat within the CNM is transported, particularly if we include cosmic ray feedback \citep{2024arXiv240307976H}. We will discuss this particular piece of physics in the later papers.

As a conclusion, in this paper we present a viable scenario for unstable phase to be dynamically stable compared to the dynamical time of radiative cooling and cloud evolution time, which can potentially explain the large unstable fraction as observed in \cite{2018ApJS..238...14M}. 
In short:
\begin{enumerate}
\item The fraction of UNM is maintained decades ($>20Myr$) over the cooling time ($\sim 2Myr)$. (Fig.\ref{fig:v_frac}) when turbulence is present. 
\item When we increase the turbulence levels, more fluid parcels fall into the thermally unstable phase. Analytically, the mass fraction of UNM is $\propto v^{\frac{2n}{1+n}}$ (Eq.\ref{eq:prediction})
\item In particular, in our numerical simulations we observe that Eq.\ref{eq:prediction} is a precise prediction of mass fraction of UNM (Fig.\ref{fig:v_frac_scaling}).
\end{enumerate}

\noindent
{\bf Acknowledgments.} KHY acknowledges Hui Li for providing extensive suggestions and comments on the conduction physics of cold filaments on the sky, particular for his insight on thermal and turbulent anisotropic conduction. KWH acknowledges Chang-Goo Kim for providing extensive suggestions and comments on the paper during the 2023 Athena++ workshop. {\toreferee We especially thank Munan Gong and Enrique Vazquez-Semadeni in providing extensive comments to our manuscript. We would also like to thank the referee for many useful comments and suggestions that significantly improved this work.} KWH \& AL acknowledge the support the NSF AST 1816234, NASA TCAN 144AAG1967, NSF grant AST 1212096 and NASA grant NNX14AJ53G.
The research presented in this article was supported by the LDRD program of LANL with project \# 20220107DR (KWH) \& 20220700PRD1 (KHY), and a U.S. DOE Fusion Energy Science project.
This research used resources from the LANL Institutional Computing Program (y23\_filaments, y24\_cmb), supported by the DOE NNSA Contract No. 89233218CNA000001.  This research also used resources of NERSC with award numbers FES-ERCAP-m4239 (PI: KHY) and m4364 (PI: KWH). This research was supported in part by grant NSF PHY-2309135 to the Kavli Institute for Theoretical Physics (KITP).   \linebreak

\noindent
{\bf Software} Julia-v1.8.2/ Julia-v1.8.3, Jupyter/miniconda3, C++ 11, MHDFlows \citep{MHDFlows}
\bibliography{sample63} 

\appendix
{\toreferee
\section{Phase diagram with and without magnetic field}
\label{ap:A}

In the main text, our description is mainly on the case of hydrodynamic turbulence. It is however unlikely that the interstellar turbulence is not magnetized. In this section, we explore how the magnetic field effect changes the result. From a theoretical point of view, the phase diagram for hydrodynamic case should smoothly transition to weakly magnetized case, noting that magnetic field does not play a role in either the pressure term nor the density term directly. An indirect effect coming from magnetic field is to act as an additional pressure support during instability. We considered this effect in \cite{instability} from a force balance perspective. However, we did not consider how the presence of magnetic field will affect the shape of the phase diagram.

Fig.\ref{fig:ap_p1} shows the phase diagrams for three sets of simulations carrying different magnetic field strength. From the left: pure hydrodynamic case without magnetic field solver turn on, magnetized case with very weak B-field ($\delta B/B=15.0$), and slightly stronger B-field ($\delta B/B=3.1$). In the case of hydrodynamic case, we observe that the phase diagram deviates from the equilibrium curve strongly and exhibits an almost {\it isobaric} transition in the unstable phase. Including the magnetic field, We observe from these simulations that a small magnetic field is sufficient to make the phase diagram in the unstable phase much steeper and closer to the equilibrium. The steepening of the phase diagram appears not dependent on the quantitative strength of magnetic field, but apparently depends on whether magnetic field exists or not. Moreover, when $\delta B/B=3.1$, we observe that the equilibrium curve accurately represents the trend of phase diagram particularly for the majority of the pixels in our simulations. 

While early studies of thermal instability with turbulence primarily focused on ideal hydrodynamical regimes (\citealt{2002ApJ...569L.127K,2007ApJ...654..945B}; See \citealt{2000ApJ...540..271V} for MHD case), the interstellar medium is both magnetized and highly turbulent in nature. In this environment, the non-linear small-scale dynamo mechanism provides a rapid amplification pathway for magnetic pressure support \citep{2021ApJ...910L..15G}.
Therefore, some modification is needed when magnetic field is included. We can start from a perspective where a portion of free kinetic energy (See discussion on kinetically driven vs magnetically driven turbulence in \citealt{2023arXiv231205399L}) is injected from large scale, and assuming that the conversion rate between solenoidal and compressible model are fix \footnote{The conversion rate is negligible in the case of incompressible, solenoidally driven turbulence, see \cite{CL03}. However, recently our team \cite{2024AAS...24411402B} found that thermal instability can convert solenoidally driven turbulence into compressible modes very efficiently, up to $\sim 50\%$. Here, we assume that a certain conversion rate is provided.} Noticing that only the compressible part of the energy will become density perturbation, while the incompressible part of the energy will be partitioned into magnetic field support. During thermal instability, the total support from {\bf both turbulence and magnetic pressure}: $ \rho v^2/l + \delta B^2/8\pi l$ is withstanding the inward collapsing thermal instability \footnote{The reason why we take the perturbed term $\delta B^2$ instead of cross term $2B\delta B$ is that the former term is larger in our set of simulations. However when the simulations are sub-Alfvenic (or be super-Alfvenic but in  sufficiently small scales, see \citealt{Lazarian06}), the cross term provides more support than the perturbed term. }. Similar to Eq.\ref{eq:10}, we write the energy flux:
\begin{equation}
\frac{2\rho v^3}{l} + \frac{\delta B^2v}{4\pi l}\sim  \frac{C\rho^{-n}v}{l} 
\end{equation}
Noticing that the scaling of magnetic field and velocity fluctuations are usually the same for all three MHD modes (see, e.g. \citealt{CL03,2010ApJ...720..742K,2020PhRvX..10c1021M}), We use $\delta B \sim v \sim l^{1/3}$ as in the main text, but we write explicitly $v \sim v_{inj} (l/L_{inj})^{1/3}$, $\delta B = \sqrt{4\pi\rho} v_A (l/L_{inj})^{1/3}$, $C=P_0/\rho_0^{-n}$, we have:
\begin{equation}
\rho v_{inj}^3 \left(\frac{l}{L_{inj}}\right) + \rho v_A^2 v_{inj}\left(\frac{l}{L_{inj}}\right) \sim \left(\frac{P_0}{\rho_0^{-n}}\right) \rho^{-n} \left(\frac{l}{L_{inj}}\right)^{1/3}
\end{equation}
which can be simplified as
\begin{equation}
\rho^{1+n} = \frac{\left(\frac{P_0}{\rho_0^{-n}}\right)}{v_{inj}(v_{inj}^2+v_A^2)}\left(\frac{l}{L_{inj}}\right)^{-2/3}
\end{equation}
where we arrive with the same scaling relation as Eq.\ref{eq:11}, but with a larger denominator. Noticing that the presence of magnetic field adds additional complexity {\it if the scaling relation of magnetic field are different from that of velocity}. This regime, however, is not consistent with observations \citep{2010ApJ...710..853C,spectrum} nor simulations \citep{2017NJPh...19f5003K,2018PhRvL.121b1104K,VDA,instability}.

\begin{figure*}[h]
\includegraphics[width=0.99\textwidth]{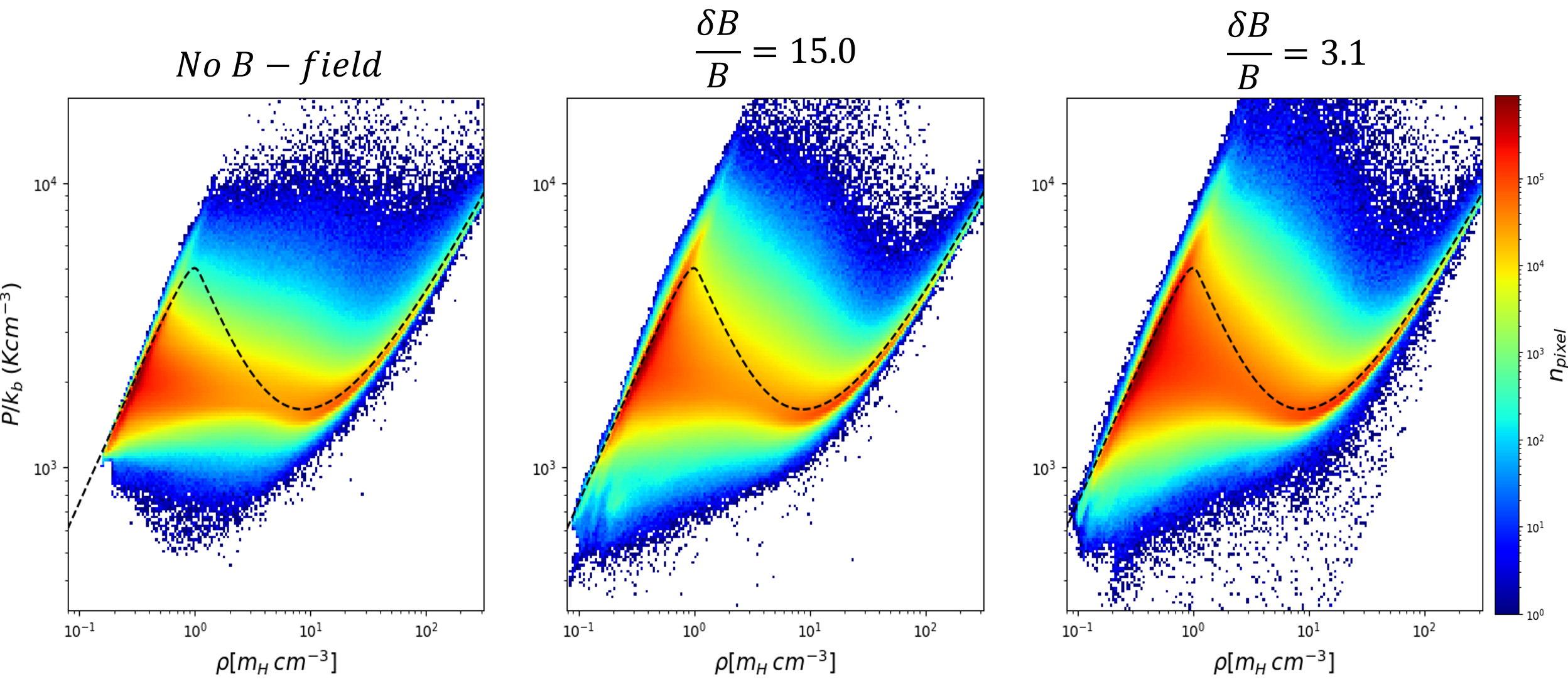}
\caption{\label{fig:ap_p1} Three panels showing how the phase diagram be different if we include different strength of mean magnetic field while keeping all other simulation parameters constant. From the left: No magnetic field (where we completely turn off the induction equation solver), a super-Alfvenic case with $\delta B/B=15.0$, and $\delta B/B=3.1$. The black curve denotes the equilibrium curve. }
\end{figure*}


\section{Distribution function approach}
\label{ap:B}

In the main text, we utilize the fact that $\rho(\bar{T}) = \int dT \rho f(T)$ and the scaling of $\rho$ in deriving the relation between velocity dispersion and gas fraction. In this section, we provide a more first-principle approach using the statistical distribution method. We notice that, while the distribution can be observed in simulations \citep{2023ApJ...946....3K} or observations (See, e.g. \citealt{2021A&A...654A..91K}), the fundamental reason on the distribution of a certain phase depends highly nonlinearly to the chemical network \citep{2023arXiv230504965G}.

First of all, we {\bf define} that the {\it volume} distribution function $f$ as a function of temperature as $f(T)$. In this case, the observables, which in this section we will mark all of them with a bar (i.e. pressure $ \bar{P} = \int dT P f(T)$, gas density $\bar{\rho}=\int dT \rho f(T)$, and temperature $ \bar{T} = \int dT T f(T)$). Notice that, $\int_{T=0}^{T=\infty} f(T) = 1$, and in principle the definition 
\begin{equation}
\bar{\rho}(\bar{T}+\overline{\Delta T}) = \frac{\int_{\bar{T}}^{\bar{T}+\overline{\Delta T}} dT \rho f(T)}{\int_{\bar{T}}^{\bar{T}+\overline{\Delta T}} dT f(T)}
\end{equation}
represents the mass fraction between $\bar{T}$ and $\bar{T}+\overline{\Delta T}$ for a small observable temperature variation $\overline{\Delta T} \ll \bar{T}$.  Our claim in the main text is that: {\bf if the scaling law is obeyed in the moment variable $\rho$, so as the observable $\bar{\rho}(\bar{T})$.} To proceed, we use the observed $f(T)$ in simulations  $f(T)=C_f \exp(T)$ in the unstable phase \citep{2023ApJ...946....3K} for some constant $C_f$ \footnote{Readers to be aware of the notation that, in Fig. 5 
of \citep{2023ApJ...946....3K} they use $f(T) = \frac{1}{V}\frac{dV}{d\log T}$ to represent the volume fraction. In their setting, the integration is unity in their form of variables. By visual observation, $\frac{1}{V}\frac{dV}{d\log T}\propto exp(T)$. }, then for a very small $\overline{\Delta T}$, we involve the equation of state from $\rho = C_T T^{-\frac{1}{n+1}}$:
\begin{equation}
\begin{aligned}
\bar{\rho}(\bar{T}+\overline{\Delta T})\approx \bar{\rho}(\bar{T})+ \frac{d\bar{\rho}}{d\bar{T}} 
&\approx \frac{C_T\overline{\Delta T}( (\bar{T}+\overline{\Delta T})^{-\frac{1}{n+1}}e^{\bar{T}+\overline{\Delta T}} - \bar{T}^{-\frac{1}{n+1}}e^{\bar{T}})}{\overline{\Delta T}  (e^{\bar{T}+\overline{\Delta T}}-e^{\bar{T}})}\\
&\approx \frac{C_T (\bar{T}+\overline{\Delta T})^{-\frac{1}{n+1}}- \bar{T}^{-\frac{1}{n+1}}+\overline{\Delta T} \bar{T}^{-\frac{1}{n+1}}}{\overline{\Delta T}}
&\approx C_T\bar{T}^{-\frac{1}{n+1}} + \frac{d\bar{\rho}}{d\bar{T}}
\end{aligned}
\end{equation}
i.e. $\bar{\rho}(\bar{T}) =C_T \bar{T}^{-\frac{1}{n+1}}$. In other words, the equation of state obeyed by the moment variables $\rho(T)$ is inherited to the statistical variable $\bar{\rho}(\bar{T})$.

}

\end{document}